\documentclass[reprint,amsmath,amssymb,prx]{revtex4-1}
\usepackage{graphicx}
\usepackage{dcolumn}
\usepackage{bm}
\usepackage{graphicx}
\usepackage{float}
\usepackage{color}
\usepackage{hyperref}
\hypersetup{colorlinks=true,citecolor=blue,urlcolor=blue,linkcolor=blue}
\usepackage{appendix}

\begin{document}

\title{Quantum well structure of double perovskite superlattice and formation of spin-polarized two-dimensional electron gas}

\author{S. Samanta, S. B. Mishra and B. R. K. Nanda}
\email{nandab@iitm.ac.in}
\affiliation{Condensed Matter Theory and Computational Lab, Department of Physics, Indian Institute of Technology Madras, Chennai-600036, India}

\begin{abstract}
Layered oxide heterostructures are the new routes to  tailor desired electronic and magnetic phases emerging from competing interactions involving strong correlation, orbital hopping, tunnelling and lattice coupling phenomena. Here, we propose a half-metal/insulator superlattice that intrinsically forms spin-polarized two-dimensional electron gas (2DEG) following a mechanism very different from the widely reported 2DEG at the single perovskite polar interfaces. From  DFT$+U$ study on Sr$_2$FeMoO$_6$/La$_2$CoMnO$_6$ (001) superlattice, we find that a periodic quantum well is created along [001] which breaks the three-fold $t_{2g}$ degeneracy to separate the doubly degenerate $xz$ and $yz$ states from the planar $xy$ state. In the spin-down channel, the dual effect of quantum confinement and strong correlation localizes the degenerate states, whereas the dispersive $xy$ state forms the 2DEG which is  robust against perturbations to the superlattice symmetry. The spin-up channel retains the bulk insulating. Both spin polarization and orbital polarization make the superlattice ideal for spintronic and orbitronic applications. The suggested 2DEG mechanism widens the scope of fabricating next generation of oxide   heterostructures.
\end{abstract}

\maketitle

\section{Introduction}

Channelizing the electronic motion through confinement is the key to the future success of fabrication of nano-scale electronic devices \cite{Cui2001}. One of the most appropriate way of achieving it is to tailor the potential profile of electrons by constructing hetero-interfaces \cite{Mani2002}, superlattices \cite{Nadvornik} and thin films \cite{Fujimori2011}, where the confinement length is comparable to the de Broglie wavelength of the associated electron. Among the heterostructures and films, the oxide families are intriguing and exhibit novel quantum states due to collective phenomena by virtue of interplay between spin, charge, and orbital degrees of freedom \cite{Hwang2012,Dagotto2005,Chakhalian2014}.

The widely investigated insulating oxide interfaces LaAlO$_3$/SrTiO$_3$ \cite{Ohtomo2004,Nakagawa2006,Karolina2009} and LaMnO$_3$/ SrMnO$_3$ \cite{Bhattacharya2008,Nanda2008} produce 2DEG to quench the polar catastrophe that arises due to alternate stacking of positively and negatively charged layers along the La side and charge neutral layers along the Sr side. A 2DEG can also be formed by quantizing the three dimensional metallic state through a confinement potential. While examples are many in semiconducting heterostructures \cite{Zhong2015,Kjaergaard2016}, it is a rare occurrence in the family of correlated oxides. One of them is the case of SrVO$_3$ ultra thin-film (8 monolayers) deposited on Nb:SrTiO$_3$ substrate \cite{Fujimori2011}. Here,  the three dimensional metallic V-$t_{2g}$ states are confined by potential well which is formed due to the Schottky barrier created at the Nb:SrTiO$_3$/SrVO$_3$ interface and the natural barrier at the SrVO$_3$/vacuum interface. Such orbital selective quantization by exploiting the $d$ orbital anisotropy forms the basic premise for the evolving area of orbitronics \cite{Tokura2003}, where the electric currents are controlled through $d$ orbital states \cite{Tokura2000}. The natural extension of orbitronics is to spin-polarize the pre-existed conducting electrons by exploiting the spin anisotropy which is one of the primary intents in this work.

To begin with, it is essential to have a source of spin-polarized conducting electrons and in this context, the double perovskite Sr$_2$FeMoO$_6$ (SFMO) has already been well established as a half-metallic system with high Curie temperature (T$_C$ $\sim$ 450 K) \cite{Kobayashi1998} and spin-polarization as large as 70\% \cite{Raghava2009}. The dispersive Mo-4$d$ ($xy$, $xz$, and $yz$) states are partially occupied in the spin-down channel, while a large band gap exists in the spin-up channel to create a  half-metallic system where the electrons are mobile in all the three dimensions. A quantum well structure can be designed to quantize the SFMO mobile electrons, by tailoring a bicolor superlattice with the other constituent being an insulator. The rare ferromagnetic insulator La$_2$CoMnO$_6$ (LCMO) is an excellent choice as its T$_C$ is close to 230 K \cite{Goodenough2003, Bull2003} and it offers a minor in-plane lattice mismatch ($\sim$ 1.5 \%) when the superlattice is grown along [001] direction.

Recent advances in modern state-of-the-art techniques such as molecular beam epitaxy and atomic layer deposition methods have paved the way to create such layered oxide superlattices. Stable nanometer thick SFMO and LCMO films, grown using PLD and RF magnetron sputtering techniques, have already been reported in the literature \cite{Du2013,Hauser2011,Galceran2016,Iliev2007}. Also experimentally, successful attempts have been made to grow free standing oxide films (e.g. VO$_2$ \cite{Li2016}, Fe$_3$O$_4$ \cite{Wu2016}, Pb(Zr,Ti)O$_3$ \cite{Jiang2017} and interfaces BiFeO$_3$/CoFe$_2$O$_4$ \cite{Amrillah2017}) using van der Waal heteroepitaxy techniques.

\begin{center}
\begin{figure}[H]
\includegraphics[angle=-0,origin=c,height=8.0cm,width=18.0cm]{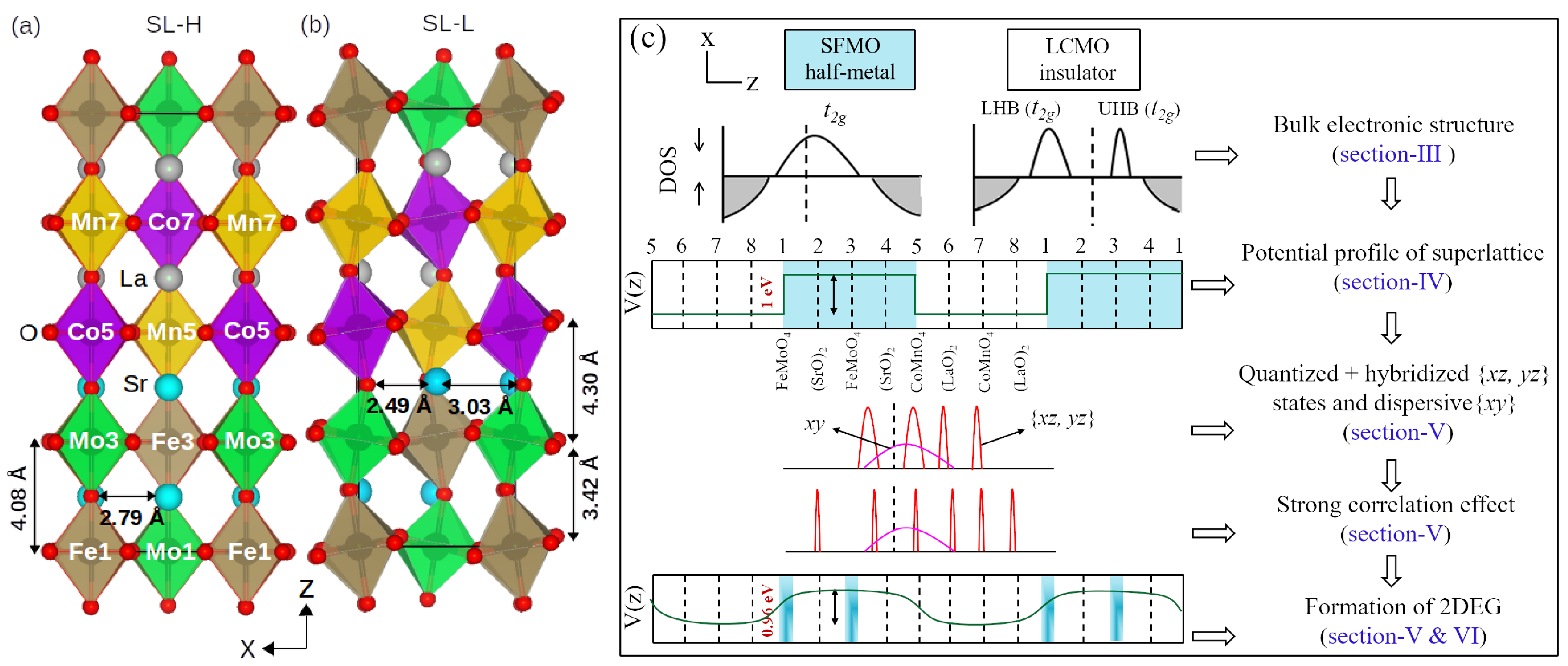}
\caption{Crystal structure of (SFMO)$_2$/(LCMO)$_2$ superlattice and mechanism for formation of 2DEG. (a) The (SFMO)$_2$/(LCMO)$_2$ superlattice assuming high symmetric SFMO as the substrate (SL-H). (b) The same superlattice, but with lower symmetric LCMO as the substrate (SL-L). The lowering in the symmetry of SL-L superlattice is due to the tilting and rotation of the octahedral complexes. (c) Schematic illustration of quantum confinement and formation of spin-polarized two dimensional electron gas in the superlattices. Here, the up-arrow and down-arrow represent the spin-up and spin-down DOS respectively and $t_{2g}$ denotes the triply degenerate ($xy$, $yz$, $xz$) states. In SFMO, it carries the character of Fe and Mo-$d$ states signifying stronger Fe-Mo hybridization. In LCMO, the strong correlation effect splits the Co-$t_{2g}$ states to form two subbands, namely, lower Hubbard band (LHB) and upper Hubbard band (UHB). The potential well of the superlattice breaks the three-fold degeneracy to two-fold degenerate $xz$ and $yz$ states and non-degenerate $xy$ state. While the former are quantized, the latter remained dispersive as in the bulk to form the 2DEG.}
\label{fig:Structure_SL_H_L_mechanism}
\end{figure}
\end{center}

In this work, we examine the (SFMO)$_2$/(LCMO)$_2$ superlattices in two different configurations as shown in Fig. \ref{fig:Structure_SL_H_L_mechanism} using DFT+$U$ calculations. In the first case, assuming the higher symmetric tetragonal SFMO as the substrate, the in-plane symmetry of the superlattice is taken to be the same as that of SFMO and we define the structure as SL-H. The second configuration (SL-L) carries the in-plane symmetry of lower symmetric monoclinic LCMO. In both the configurations, the atomic positions are relaxed to obtain the ground state. 

The electronic and magnetic ground state of both SL-H and SL-L superlattices reveal no new magnetic ordering. However, a periodic quantum well with depth close to 1 eV is  developed along the z-axis (growth direction) due to the difference in the chemical potential of the constituents. As a consequence, there is an orbital selective quantization of the fractionally occupied $t_{2g}$ itinerant states. The strong correlation further localizes these quantized states. During the whole process the planar $xy$ dispersive state remains unchanged which leads to the evolution of a two dimensional spin-polarized electron gas (2D-SPEG) from a 3D-SPEG. The mechanism, as understood from the electronic structure calculations presented in this paper, is schematically illustrated in Fig. \ref{fig:Structure_SL_H_L_mechanism}. It completely differs from the mechanism of charge reconstruction involved in single perovskite polar interfaces and hence opens up new avenues to synthesize next generation heterostructures out of non-polar correlated oxides in order to create 2DEG for practical purposes.

\onecolumngrid
\twocolumngrid

\section{Computational details}

Density functional theory (DFT) calculations are carried out using both pseudopotential method with plane wave basis set as implemented in  Quantum espresso simulation package \cite{Paolo2009} and full potential linearized augmented plane wave method with the basis set includes local orbitals (FP-LAPW+lo) as implemented in WIEN2k simulation package \cite{Blaha2001}. For both the cases the PBE-GGA \cite{Perdew1996} exchange correlation functional is considered and an $8\times8\times8$ Monkhorst-Pack k-mesh is used for the Brillouin Zone integration.

The pseudopotential method is used for structural optimization  and for the calculation of the electrostatic potential  in the real space. For structural relaxation, the kinetic energy cutoff for the planewaves is set to 30 Ry. The electron-ion interaction is considered within Vanderbilt ultrasoft pseudopotential for which charge density cutoff is chosen to be 300 Ry. The tolerance for the Hellmann-Feynman force on each atom is taken as 20 meV/\AA.

The optimized structure obtained from pseudopotential method, has been further used for the calculation of electronic and magnetic structure using FP-LAPW method. The computational details for this method are as follows. To incorporate the effect of strong correlation, an effective onsite correlation parameter $U$ ($U_{eff}$ = $U-J$) is included through rotationally invariant Dudarev approach \cite{Dudarev1998}. All the results in the paper are presented for $U$ = 3 eV. However, to examine the invariance of the mechanism, the results are also analyzed for higher $U$ (= 5 eV). The LAPW basis function considers 5$d$ and 6$s$ of La; 5$s$ of Sr; 3$d$ and 4$s$ of Mn, Fe, and Co; 4$d$ and 4$s$ of Mo, and 2$s$ and 2$p$ of O. The $RK_{max}$ is taken to be 7.0 yielding 24235 plane waves for each k-point in the interstitial region. The principal component of conductivity tensors ($\sigma_{\alpha\beta}$) are computed using semi-classical Boltzmann transport theory as implemented in BoltzTraP code \cite{Madsen2006}. A highly dense non-shifted mesh with 32000 k-points is used to obtain the smooth interpolation of bands and to compute the necessary derivatives which is required for the calculation of $\sigma_{\alpha\beta}$.\\

\section{Bulk Electronic Structure}

\begin{center}
\begin{figure}[H]\hspace*{1cm}
\includegraphics[angle=0,origin=c,height=6.0cm,width=16.0cm]{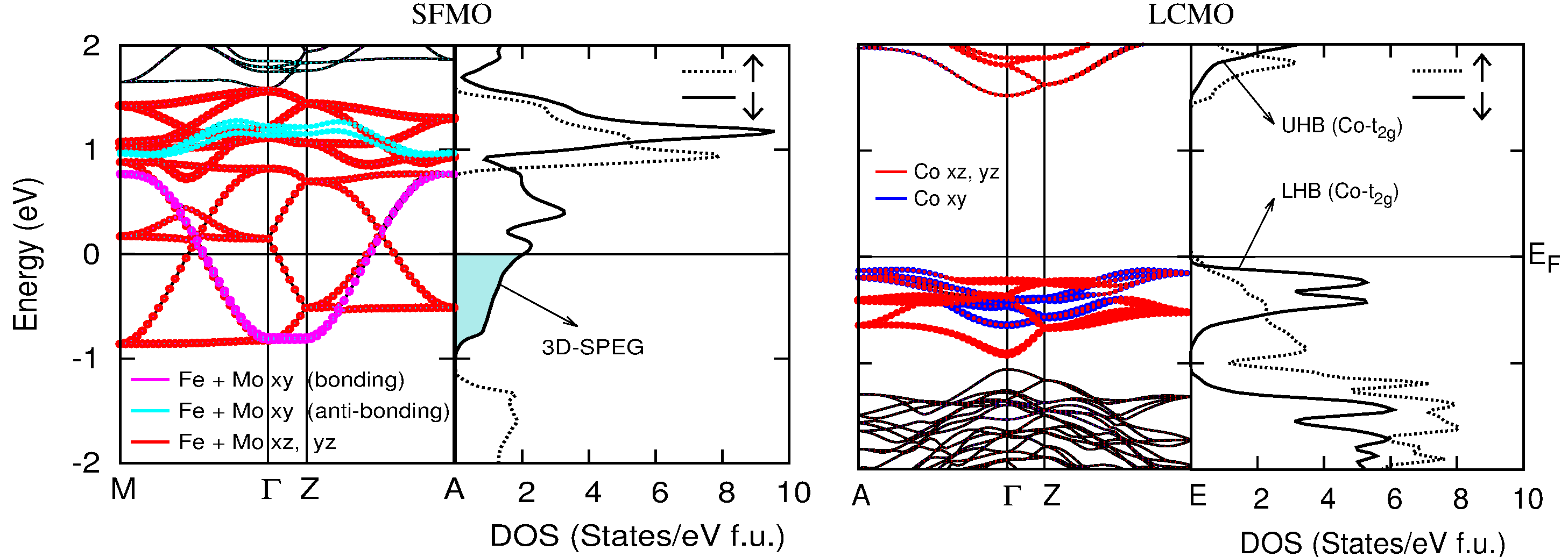}
\caption{Band structure (shown in spin-down channel) and densities of states for SFMO and LCMO. The results are obtained using GGA+$U$ ($U$ = 3 eV). A four formula unit cell is used in order to have an identical Brillouin zone as that of the superlattice. The partially occupied $t_{2g}$ states in the spin-down channel form the 3DEG. LCMO exhibits insulating ground state following Mott mechanism.}
\label{fig:bulk_ES_GS_mechanism}
\end{figure}
\end{center}

Bulk SFMO is a half-metallic ferrimagnet, where only the spin-down channel exhibits the metallic behavior and Fe spins are aligned antiparallel to Mo spins \cite{Kobayashi1998,Sarma2000}. As Fig. \ref{fig:bulk_ES_GS_mechanism} shows, the Fermi level (E$_F$) in the spin-down channel is occupied by the Mo predominant bonding states of the Mo-$t_{2g}\downarrow$ - Fe-$t_{2g}\downarrow$ hybridization. In the $d^5$ configuration of the high-spin Fe$^{3+}$ ion, the $t_{2g}\downarrow$ state is expected to be empty. Similarly, in the $d^1$ configuration of Mo$^{5+}$ ion, the $t_{2g}\downarrow$ state is partially occupied while the   $d\uparrow$ states are empty. However, the delocalized 4$d$ states hybridize significantly with the Fe-$t_{2g}\downarrow$ states to form a set of partially occupied dispersive bands. As a consequence, a three dimensional spin-polarized electron gas is formed.

\onecolumngrid
\twocolumngrid

Double perovskite LCMO is a rare ferromagnetic insulator \cite{Baidy2011}. Its band structure and densities of states (DOS) are shown in Fig. \ref{fig:bulk_ES_GS_mechanism}. While the Mn stabilizes in 4+ charge state, leading to $t_{2g}^3\uparrow$ $e_{g}^0\uparrow$ $t_{2g}^0\downarrow$ $e_{g}^0\downarrow$ configuration, the Co stabilizes in 2+ charge state leading to $t_{2g}^3\uparrow$ $e_{g}^2\uparrow$ $t_{2g}^2\downarrow$ $e_{g}^0\downarrow$ configuration. In the spin-up channel, the band gap is opened by the large crystal field split of Mn-$d$ state as well as large spin-exchange split of both Mn and Co-$d$ states \cite{Zhu2012}. In the absence of strong correlation effect, $t_{2g}^2\downarrow$ configuration would have created a metallic state for the perfect cubic phase. However, with tilting of the octahedra as well as strong correlation effect, the $t_{2g}$ states are further split into occupied lower Hubbard band and unoccupied upper Hubbard band to open up a gap in the spin-down channel to make the system insulating \cite{Parida2017}. Our estimated exchange energies ($J$ = $E_{\uparrow\downarrow}-E_{\uparrow\uparrow}$) confirm that there is a strong ferromagnetic coupling between the Co and Mn spins ($J_{Co-Mn}\sim$ 10.11 meV) \cite{Lv2012} which overcomes the Co-Co ($J_{Co-Co}\sim-1.92$ meV) and Mn-Mn ($J_{Mn-Mn}\sim-1.52$ meV) antiferromagnetic couplings. The detailed mechanisms are illustrated in the appendix to further elucidate the half-metallic and ferromagnetic-insulating behavior of SFMO and LCMO respectively.

\section{Formation of Periodic Quantum Well Structure} 

The growth of the (SFMO)$_2$/(LCMO)$_2$ superlattice, as shown in Fig. \ref{fig:Structure_SL_H_L_mechanism}, brings a potential mismatch between the SFMO and LCMO site and hence, creates a quantum well structure. To demonstrate it, we have estimated the variation of the macroscopic average of electrostatic potential ($V^{MA}$) of bulk SFMO and LCMO as well as that of the SL-H superlattice as follows. First, the $xy$-planar average of the potential ($V^{PA}$) is obtained by averaging the raw three dimensional potential ($V^{raw}$) \cite{Franciosi1996}.

\begin{equation}
V^{PA}(z)=\frac{1}{S}\int_sV^{raw}(x,y,z)dxdy.\\
\end{equation}

Where $S$ is the area of the (001) plane of the unit cell. The $V^{PA}$ is further averaged to obtain $V^{MA}$.

\begin{equation}
V^{MA}(z)=\frac{1}{c}\int^{z+c/2}_{z-c/2}V^{PA}(z^\prime)dz^\prime\\
\end{equation}

Here, $c$ is the length of one period. For LCMO and SFMO slabs, the respective lattice parameters are taken as $c$. In the case of superlattice, the V$^{MA}$ is calculated using the $c$ lattice parameter of SFMO as well as that of LCMO and the average of the two is considered to minimize the error at the interface \cite{Koberidze2016}.

\begin{figure}[H]
\begin{center}
\includegraphics[angle=0,origin=c,height=9.2cm,width=8.5cm]{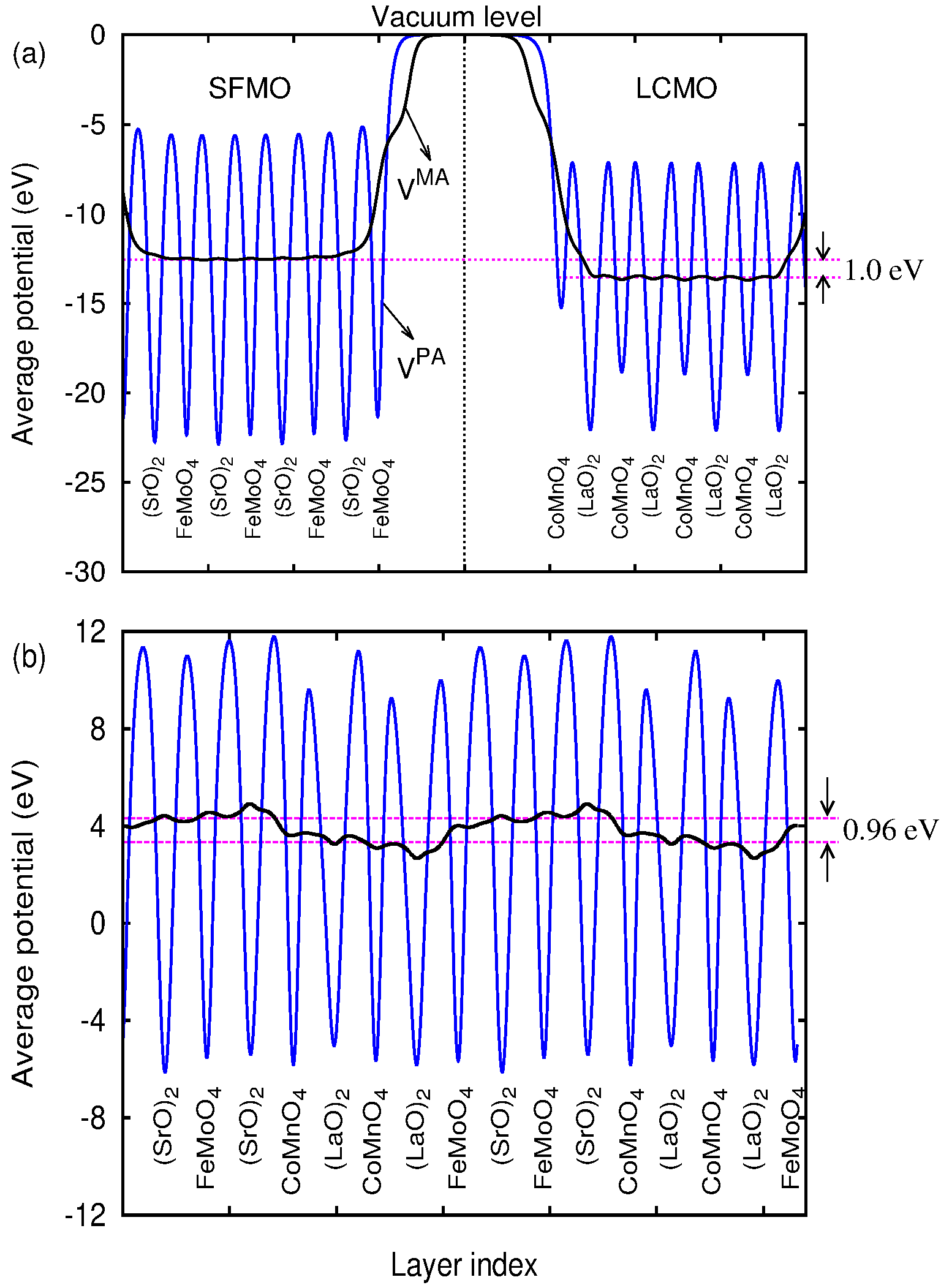}
\caption{(a) Planar average (V$^{PA}$) and macroscopic average (V$^{MA}$) potential of 4 unit cell thick SFMO and LCMO slabs  with reference to the vacuum level. (b) V$^{PA}$ and V$^{MA}$ of the SL-H superlattice suggesting the formation of periodic quantum well structure.}
\label{fig:avg_potential}
\end{center}
\end{figure}

Fig. \ref{fig:avg_potential} shows the V$^{PA}$ and V$^{MA}$ for the pure LCMO and SFMO slabs as well as for the SL-H superlattice. When compared with the vacuum level, the $V^{MA}$ of the SFMO slab is found to be $\sim$ 1 eV higher than that of the LCMO side. Hence, in the absence of any significant ionic displacements and breakdown of the planar geometry, the SL-H superlattice is expected to produce a periodic quantum well structure with depth of 1 eV. Our structural relaxation on the SL-H superlattice suggests that non-planar displacement of the ions are of the order of 0.004 {\AA} and therefore, the layered geometry is maintained. Also, Fig. \ref{fig:avg_potential} (b) infers that the $V^{MA}$ of the superlattice in the ground state structure shows a periodic quantum well of depth $\sim$ 0.96 eV.

The spin-polarized 3DEG of the superlattice will now experience this periodic quantum well and also, the strong correlation effect. Hence, new quantum states are expected to emerge which we have examined by carrying out band structure calculations.

\section{Eigenstates reconstruction and formation of 2DEG}

In the spin-up channel, bulk SFMO and LCMO exhibit a band gap larger than the depth of the potential well (see Fig. \ref{fig:bulk_ES_GS_mechanism}). Therefore, in this spin channel, like the bulk, the superlattices also exhibit insulating behavior. Hence, our band structure analysis for the superlattice is restricted to the spin-down channel.

\begin{center}
\begin{figure}[H]\hspace*{2cm}
\includegraphics[angle=-0,origin=c,height=4.7cm,width=12.0cm]{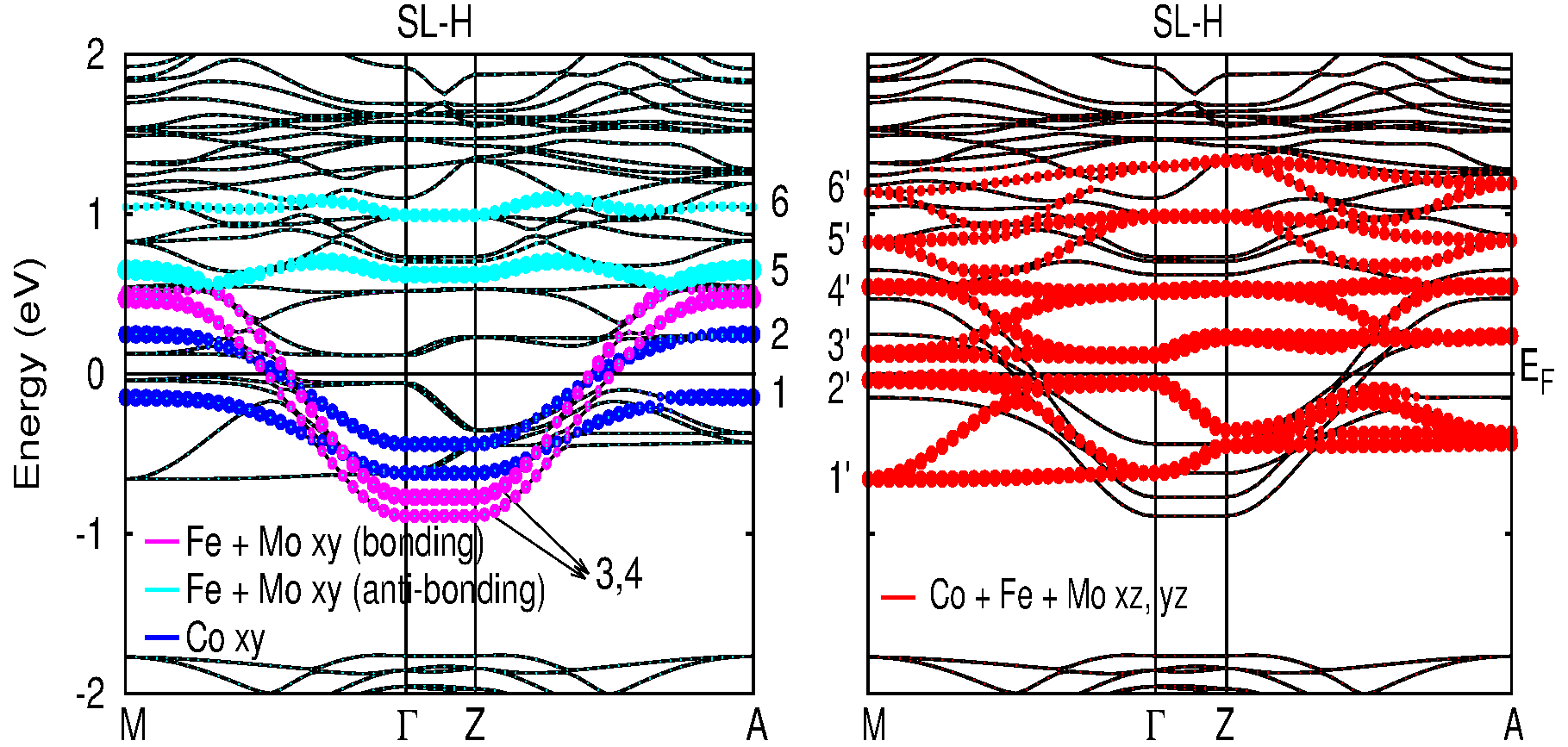}
\caption{Spin-down band structure of SL-H superlattice for $U$ = 0. The contribution of the planar orbitals (Mo, Fe, and Co -$xy$) and the $z$ axis oriented orbitals ($xz$ and $yz$) to the band structure are shown on the left and the right respectively. The discrete pairs 1$^{\prime}$ to 6$^{\prime}$ are the outcome of the quantization through the periodic potential well (see Fig. \ref{fig:avg_potential}). The partially occupied bands, 3 and 4 of the left panel, forms the spin-polarized 2DEG.}
\label{fig:SL-H_u0}
\end{figure}
\end{center}

The $t_{2g}$ projected spin-down band structure of the SL-H superlattice within the independent electron approximation ($U$ = 0) is shown in Fig. \ref{fig:SL-H_u0}. Since Mn-$t_{2g}\downarrow$ states lie far above the E$_F$ due to large exchange splitting, the effect of the potential well is inconsequential. For the remaining six transition metal elements (two of each Fe, Mo, and Co), the periodic potential well along $z$ breaks the three-fold degeneracy and splits the corresponding $t_{2g}\downarrow$ states into planar $xy$ and two-fold degenerate $xz$ and $yz$ states. The left panel of Fig. \ref{fig:SL-H_u0} highlights the $xy$ orbital dominated bands. The two lower lying (nearly) occupied parabolic bands (1, 2; blue) belong to two Co atoms located in the lower potential region. Out of the remaining four, two of them are partially occupied parabolic bonding bands (3, 4; magenta) and two of them are the unoccupied anti-bonding bands (5, 6; cyan) resulted out of Fe-Mo $t_{2g}\downarrow-t_{2g}\downarrow$ interactions as discussed in the bulk band structure (Fig. \ref{fig:bulk_ES_GS_mechanism}). Except for a minor shift in their energy levels, these bonding bands resemble to that of the bulk ($U$ = 3 eV) band structure which suggests that these states are delocalized and are not affected by the quantum well.

The right panel of Fig. \ref{fig:SL-H_u0}, highlights the bands dominated by the orbitals ($xz$, $yz$). Unlike the bands with in-plane $xy$ states, these bands, lying in the range E$_F$ - 0.8 to E$_F$ + 1.2 eV, are found to be localized and discrete which is a signature of quantization. Due to degeneracy of $xz$ and $yz$ states, there are six pairs of such bands (1$^{\prime}$ to 6$^{\prime}$ of Fig. \ref{fig:SL-H_u0} (right)). The lower, middle, and upper two pairs are predominantly contributed by Co, Fe, and Mo atoms respectively. However, reasonable presence of Mo-$\{xz, yz\}$ characters in the lower two pairs suggests that a new Mo-$t_{2g}$ - Co- $t_{2g}$ hybridization has taken place across the interface.

\onecolumngrid
\twocolumngrid

\begin{center}
\begin{table}[H]
\caption{Total energy of different magnetic configurations of the superlattice. In bulk magnetic ordering, the spins of the transition metal cations in SFMO are antiparallel, whereas they are parallel in LCMO. In C-AFM configuration, the intra-plane coupling between the neighboring spins is antiferromagnetic, while inter-plane coupling is ferromagnetic. In G-AFM spin arrangement, both intra and inter-plane couplings between the neighboring spins are antiferromagnetic. The A-AFM arrangement corresponds to intra-plane ferromagnetic coupling and inter-plane antiferromagnetic coupling. We find that there are no new magnetic phases and the superlattice inherits the spin-arrangement of the respective bulk compounds.}
\vspace{0.5cm}
\label{tab:magnetic_ordering} 
\begin{tabular}{ c c c c }
\hline
\multicolumn{2}{ c }{Interface magnetic orderings} & \multicolumn{2}{ c }{\hspace{0.2cm}$\Delta$E in eV}\\
\cline{3-4}
\multicolumn{2}{ c }{} & \hspace{0.7cm}SL-H & \hspace{0.7cm} SL-L\\
\hline
Bulk & \hspace{0.3cm}$\mathrm{(Co)_\uparrow(Mn)_\uparrow/(Fe)_\uparrow(Mo)_{\downarrow}}$ & \hspace{0.7cm}0 & \hspace{0.7cm}0\\
\hline
C-AFM & \hspace{0.3cm}$\mathrm{(Co)_\uparrow(Mn)_\downarrow/(Fe)_\uparrow(Mo)_\downarrow}$ & \hspace{0.7cm}0.82 & \hspace{0.7cm}0.19\\
\hline
G-AFM & \hspace{0.3cm}$\mathrm{(Co)_\uparrow(Mn)_\downarrow/(Fe)_\downarrow(Mo)_\uparrow}$ & \hspace{0.7cm}0.73 & \hspace{0.7cm}0.20\\
\hline
A-AFM & \hspace{0.3cm}$\mathrm{(Co)_\uparrow(Mn)_\uparrow/(Fe)_\downarrow(Mo)_\downarrow}$ & \hspace{0.7cm}0.09 & \hspace{0.7cm}0.10\\
\hline
\end{tabular}
\end{table}
\end{center}

The independent electron approximation does not provide the exact ground state, particularly in the case of oxides, as there is inadequacy in accounting the electron correlation in the system. The correlation effect can be included through the parametric Hubbard $U$ formalism. As the ground state electronic structure of bulk SFMO and LCMO is accurately estimated for $U$ = 3 eV, we have considered the same for the superlattices as well. Also, to determine the ground state magnetic configuration, several possible arrangements of the Co, Mn, Fe, and Mo spins are considered and the corresponding total energies are estimated in Table \ref{tab:magnetic_ordering}. We find that there is no magnetic reconstruction and the bulk magnetic ordering of SFMO and LCMO constitutes the magnetic ground state of the superlattice.

\begin{center}
\begin{figure}[H]
\includegraphics[angle=-0,origin=c,height=9.0cm,width=18.0cm]{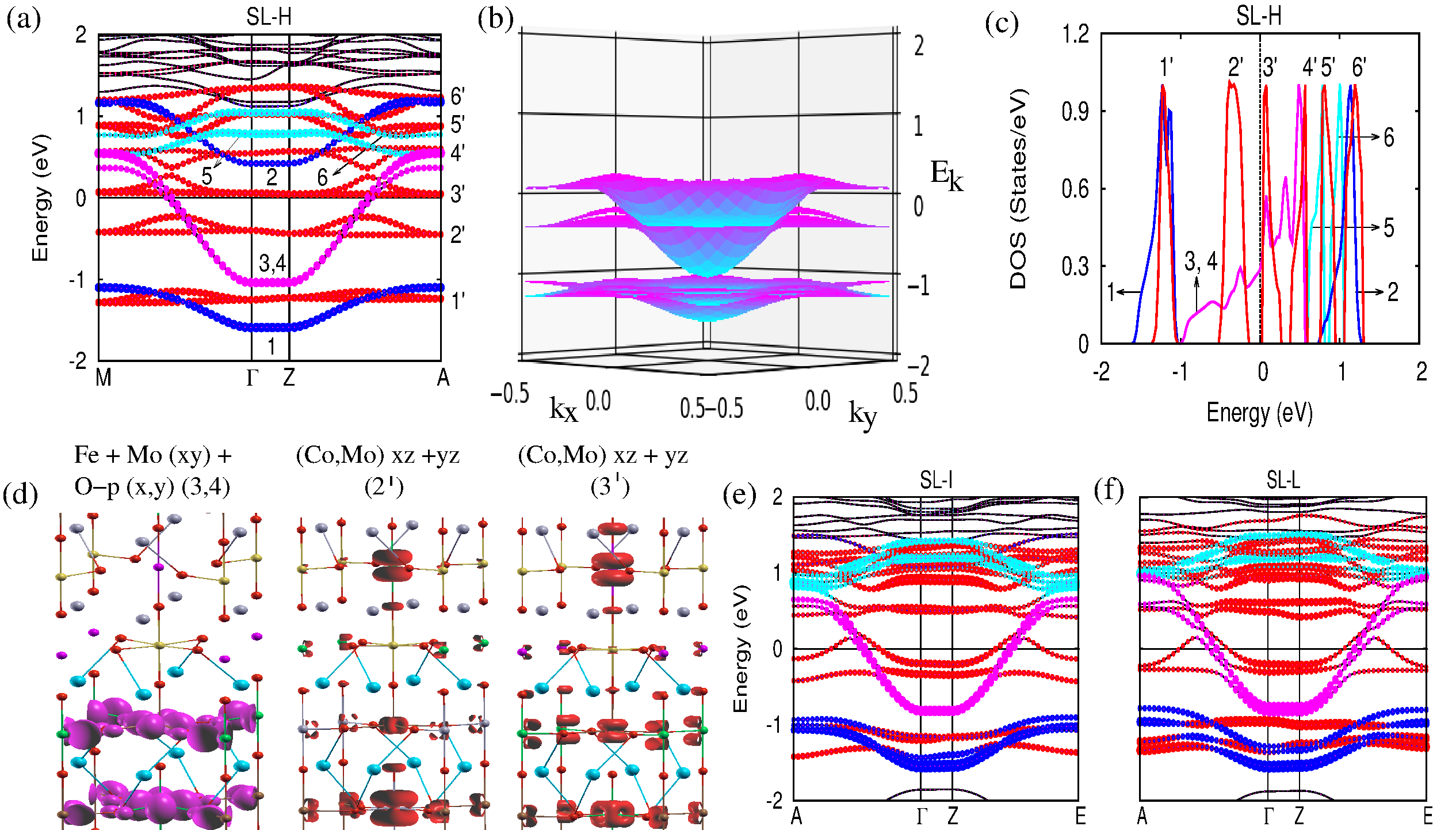}
\caption{Strongly correlated electronic structure of the SFMO/LCMO superlattice. (a) The spin-down band structure of the SL-H superlattice projecting the contribution of in-plane ($xy$) and out-of-plane ($xz, yz$) orbitals on the bands as obtained using  GGA$+U$ (= 3 eV). For the color code of the bands, refer Fig. \ref{fig:SL-H_u0}. (b) The same spin-down band structure, but now plotted in the interfacial reciprocal space ($k_x-k_y$) plane and in the vicinity of E$_F$. (c) The spin-down DOS of the SL-H superlattice. The numbering of the bands is identified with discretization due to the effect of both the periodic quantum well and strong correlation. (d) (left to right) The charge density plots for the dispersive bands $\{3, 4\}$ and quantized bands $2^{\prime}$ and $3^{\prime}$. The former further confirms that the itinerant electrons, occupied by the hybridized states, are formed by the $xy$, $x$ and $y$ orbitals of the FeMoO$_4$ plane. The charge densities also re-verify that the quantized bands $2^{\prime}$ and $3^{\prime}$ are formed by the out-of-plane $xz$ and $yz$ orbitals. (e) and (f) represent the spin-down band structure for the superlattice with intermediate symmetry (SL-I) and with lower symmetry (SL-L) superlattices (see Fig. \ref{fig:Structure_SL_H_L_mechanism} (b)) respectively. While the lowering in symmetry further discretizes the quantized bands, the partially occupied dispersive bands $\{3, 4\}$ are almost unaffected.}
\label{fig:SL_H_I_L_ES_MLWF}
\end{figure}
\end{center}

The spin-down band structure for the magnetic ground state of SL-H superlattice is shown in Fig. \ref{fig:SL_H_I_L_ES_MLWF}(a). We find that following the Mott mechanism, there is a significant re-positioning of the bands w.r.t. the $U$ = 0 band structure. Out of the two lower lying Co-$xy$ dominated bands (1 and 2 of Fig. \ref{fig:SL-H_u0} (left)), the occupied one lowers its energy by roughly 1 eV and the fractionally occupied one raises its energy approximately by 1 eV to become unoccupied. However, the fractionally occupied itinerant Mo-Fe bonding $xy$ states remain unchanged. Similarly, in the case of potential well quantized bands, dominated by $xz$, $yz$ characters, the lowest pair (1$^{\prime}$) is pushed down further below and lies at $-$1.2 eV w.r.t. E$_F$. Also, there is a visible separation of 0.5 eV between the next two quantized pairs (2$^{\prime}$ and 3$^{\prime}$). While 2$^{\prime}$ is completely occupied, 3$^{\prime}$ is empty. These two quantized states are now dominated with Mo and Co-{$xz$, $yz$} characters. The upper three quantized states are less affected by the $U$ effect. In addition to the re-positioning, the strong correlation effect further localizes the quantized states. The band dispersion, plotted in the interfacial reciprocal space ($k_x-k_y$ plane) (Fig. \ref{fig:SL_H_I_L_ES_MLWF} (b)) and the eigenstate resolved DOS for the whole Brillouin zone (Fig. \ref{fig:SL_H_I_L_ES_MLWF} (c)), further confirm the quantization and localization of the $xz$ and $yz$ based bulk states and the presence of unaffected itinerant bonding $xy$ states. The charge density plot of Fig. \ref{fig:SL_H_I_L_ES_MLWF} (d) provides a visualization assessment of the orbital contribution of bands 3, 4, 2$^{\prime}$, and 3$^{\prime}$.

\onecolumngrid
\twocolumngrid

To examine the robustness of the quantization and the 2DEG, we have examined the ground state electronic structure of the lower symmetry structure (SL-L, Fig \ref{fig:Structure_SL_H_L_mechanism}(b)), where the SL is designed assuming the SL is grown on a LCMO substrate. Also, the electronic structure of an intermediate symmetry (SL-I), designed by taking the average of LCMO and SFMO crystal, is calculated. Figs. \ref{fig:SL_H_I_L_ES_MLWF} (e and f) show the spin-down band structure of SL-I and SL-L configurations. While the crystal structure of SL-H is tetragonal, it is monoclinic for SL-I and SL-L.  With lowering in the symmetry from tetragonal to monoclinic ($\beta \neq 90^0$,  there is an intermixing of the $xy$ state with the $xz$ and $yz$ states through inter-site hybridization. This results in discretization and minor localization of the planar $xy$ dominated bands as can be observed from Figs. \ref{fig:SL_H_I_L_ES_MLWF} (e and f). However, the effect is very weak and can be neglected. Later from Fig. \ref{fig:conductivity}, we will find that the electrical conductivity along the superlattice growth direction is negligible for all the superlattices confirming the electron conduction confined to the $xy$ plane.

\begin{figure}[H]
\begin{center}
\includegraphics[angle=-0,origin=c,height=6.0cm,width=8.0cm]{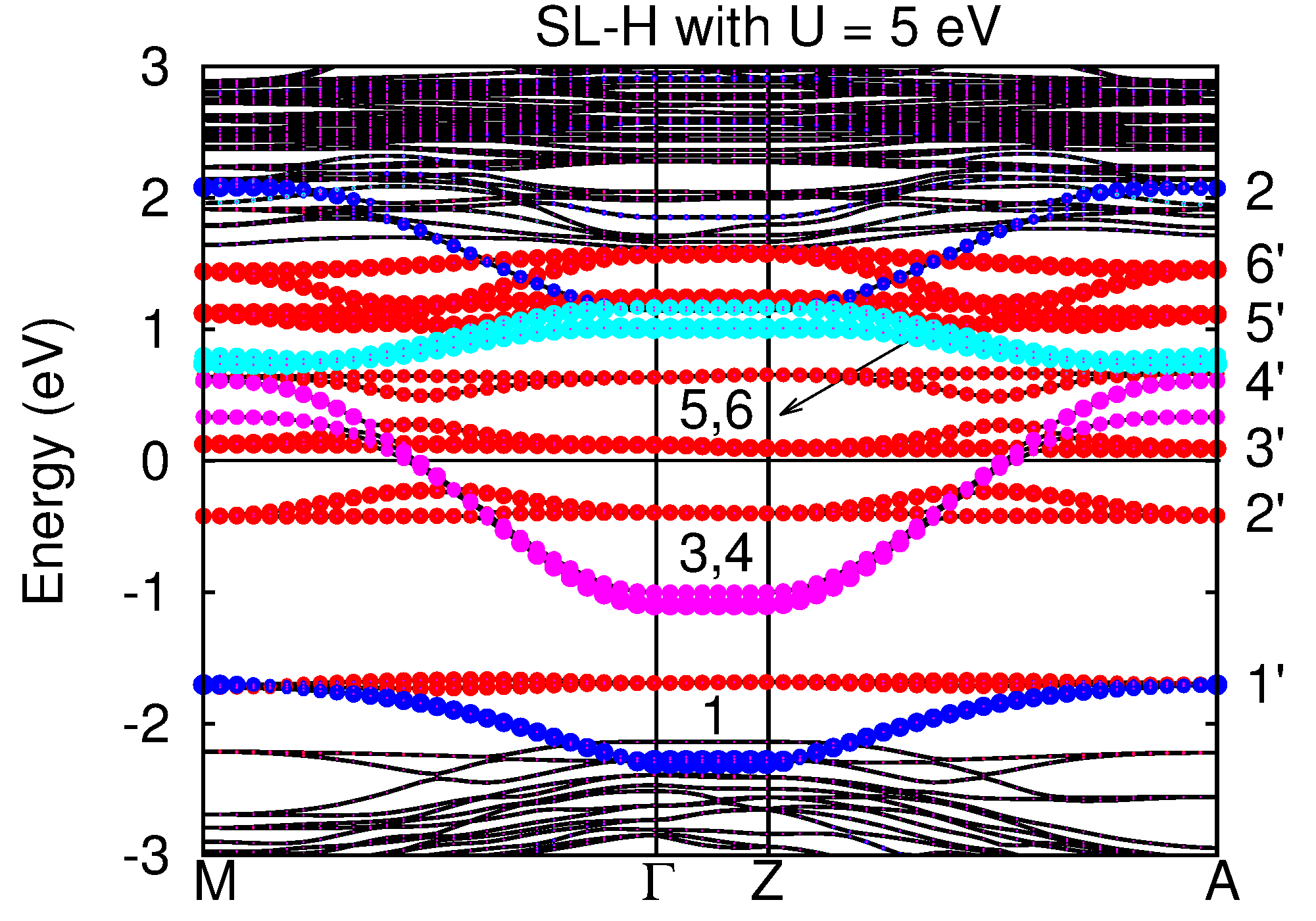}
\caption{The planar $xy$ projected bands (1 to 6) and $z$-axis oriented $xz$ and $yz$ orbitals projected bands (1$^{\prime}$ to $6^{\prime}$) in the spin-down channel for the SL-H superlattice. The results are obtained for $U$ = 5 eV.}
\label{fig:SL_H_u5}
\end{center}
\end{figure}

To see if the spin-polarized 2DEG and the quantized states remain invariant with respect to strong correlation effect, we have further examined the electronic structure for higher values of $U$. The spin-down band structure of SL-H superlattice for $U$ = 5 eV is shown in Fig. \ref{fig:SL_H_u5}. We find that the formation of 2DEG is invariant. However, higher value of $U$ further localizes the quantized states. In Appendix-C, we have applied different $U$ to different transition elements and still find that it has no bearing over the formation of 2DEG.

\section{Transport properties}

The formation of spin-polarized 2DEG out of bulk SFMO 3DEG through confinement effect, can be quantified by calculating the conductivity. For this, we have adopted semi-classical Boltzmann transport theory as implemented in BoltzTraP code \cite{Madsen2006} and calculated the conductivity tensor ($\sigma$) from the first order derivatives of bands $\epsilon (k)$.

\begin{equation}
\sigma_{\alpha\beta}(\epsilon)=\frac{e^2\tau}{N}\sum_{i,\textbf{k}} v_\alpha(i,\textbf{k})v_\beta(i,\textbf{k})\frac{\delta(\epsilon-\epsilon_{i,\textbf{k}})}{d\epsilon},
\label{eqn:transport_eqn3}
\end{equation}

where $\tau$ is the relaxation time, $i$ is the band index, $v$ is the first order derivative of $\epsilon_{i,\textbf{k}}$, and N is the number of $\textbf{k}$ points sampled. The notations $\alpha$ and $\beta$ stand for the crystal axes. The temperature dependent conductivity as evaluated using Eq. \ref{eqn:transport_eqn3} is given below.

\begin{equation}
\sigma_{\alpha\beta}(T,\mu) =\frac{1}{\Omega}\int\sigma_{\alpha\beta}(\epsilon) \Big[-\frac{\partial f_\mu(T,\epsilon)}{\partial\epsilon} \Big]d\epsilon,
\label{eqn:transport_eqn}
\end{equation}

where $\Omega$ is the volume of the unitcell, $\mu$ (= E$_F$) is the chemical potential and $f$ is the Fermi-Dirac distribution function. 

\begin{center}
\begin{figure}[H]
\includegraphics[angle=-0,origin=c,height=5.0cm,width=18.0cm]{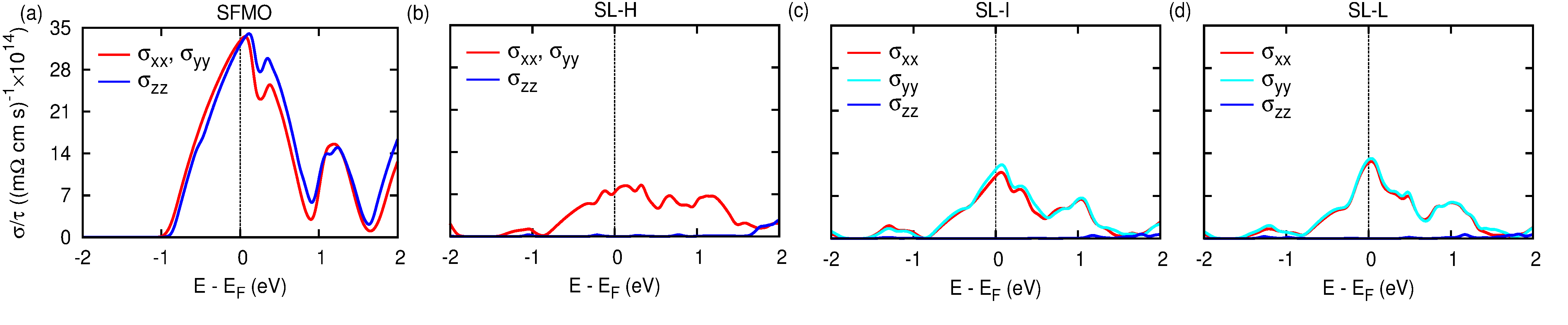}
\caption{Transport properties of bulk SFMO and SFMO/LCMO superlattice (a) - (d). The principal component of  electrical conductivity tensor at room temperature for bulk SFMO, SL-H, SL-I, and SL-L superlattices respectively. The results are obtained from Eq. \ref{eqn:transport_eqn} using semi-classical Boltzmann transport theory. The confinement potential restricts the electron motion along [001] and hence, $\sigma_{zz}$ becomes negligible. Significant values of $\sigma_{xx}$ and $\sigma_{yy}$ imply two dimensional mobility of the electrons and hence, the the formation of spin-polarized 2DEG. Due to $xy$ planar symmetry, the $\sigma_{xx}$ and $\sigma_{yy}$ are same in bulk SFMO and SL-H superlattice. Minor distortion in the plane makes them distinguishable in SL-I and SL-L superlattices.}
\label{fig:conductivity}
\end{figure}
\end{center}

In Fig \ref{fig:conductivity}, we have plotted $\sigma/\tau$ vs $E-E_F$ at room temperature for both bulk and the superlattices. In bulk SFMO, conductivity along all the three principal axes are nearly same as it has partially occupied dispersive three-fold degenerate $t_{2g}$ states (see Fig. \ref{fig:bulk_ES_GS_mechanism}). In contrast, for the SL-H superlattice, the potential along $z$ restricts the electronic motion along [001]. Hence, $\sigma_{xx}/\tau$ and $\sigma_{yy}/\tau$ are finite, but $\sigma_{zz}/\tau$ is negligible. However, the magnitude of conductivity along $x$ or $y$ has reduced, approximately by two-third, compared to the bulk. It is due to the fact that in SL-H, $xz$ and $yz$ orbitals are no longer dispersive. Only the bonding dispersive band dominated with the planar $xy$ orbital contributes to the conductivity. The $\sigma/\tau$ vs $E-E_F$ plot, for the SL-I and SL-L superlattices with reduced symmetry also shows similar conductivity phenomena as that of SL-H suggesting the robustness of the spin-polarized 2DEG against any perturbation through lattice distortion.\\

\onecolumngrid
\twocolumngrid

\vspace*{-2cm}

\section{Conclusions}

In summary, using DFT+$U$ method, we have shown that a magnetic metal-insulator superlattice Sr$_2$FeMoO$_6$/La$_2$CoMnO$_6$ creates a spin-polarized 2DEG (SP-2DEG). Our study provides an alternate quantization mechanism to intrinsically form 2DEG which is very different from the conventional engineering of polar hetero-interfaces to achieve the same. The quantization mechanism involves confinement of the spin-polarized mobile electrons through a periodic finite square well potential and further localization of the quantized states through strong correlation effect. This restricts the mobility of the electron gas to the plane perpendicular to the potential well. An experimental realization of such a superlattice will be an ideal platform to study several fundamental phenomena like intrinsic anomalous Hall effect and Rashba effect. Since the bulk magnetic order is unaffected in this superlattice, it is expected to have high Curie temperature as in the bulk. Therefore, the SP-2DEG formed here, will be useful for spintronic applications.

\vspace*{-0.4cm}

\section{Acknowledgements}

The authors would like to thank HPCE, IIT Madras for providing the computational facility. This work is supported by Department of Science and Technology, India, through Grant No. EMR/2016/003791.

\appendix

\section{Mechanism of half-metallic behavior in Sr$_2$FeMoO$_6$ and insulating behavior in La$_2$CoMnO$_6$}

\begin{figure}[H]
\begin{center}
\includegraphics[angle=-0,origin=c,height=7.4cm,width=8.8cm]{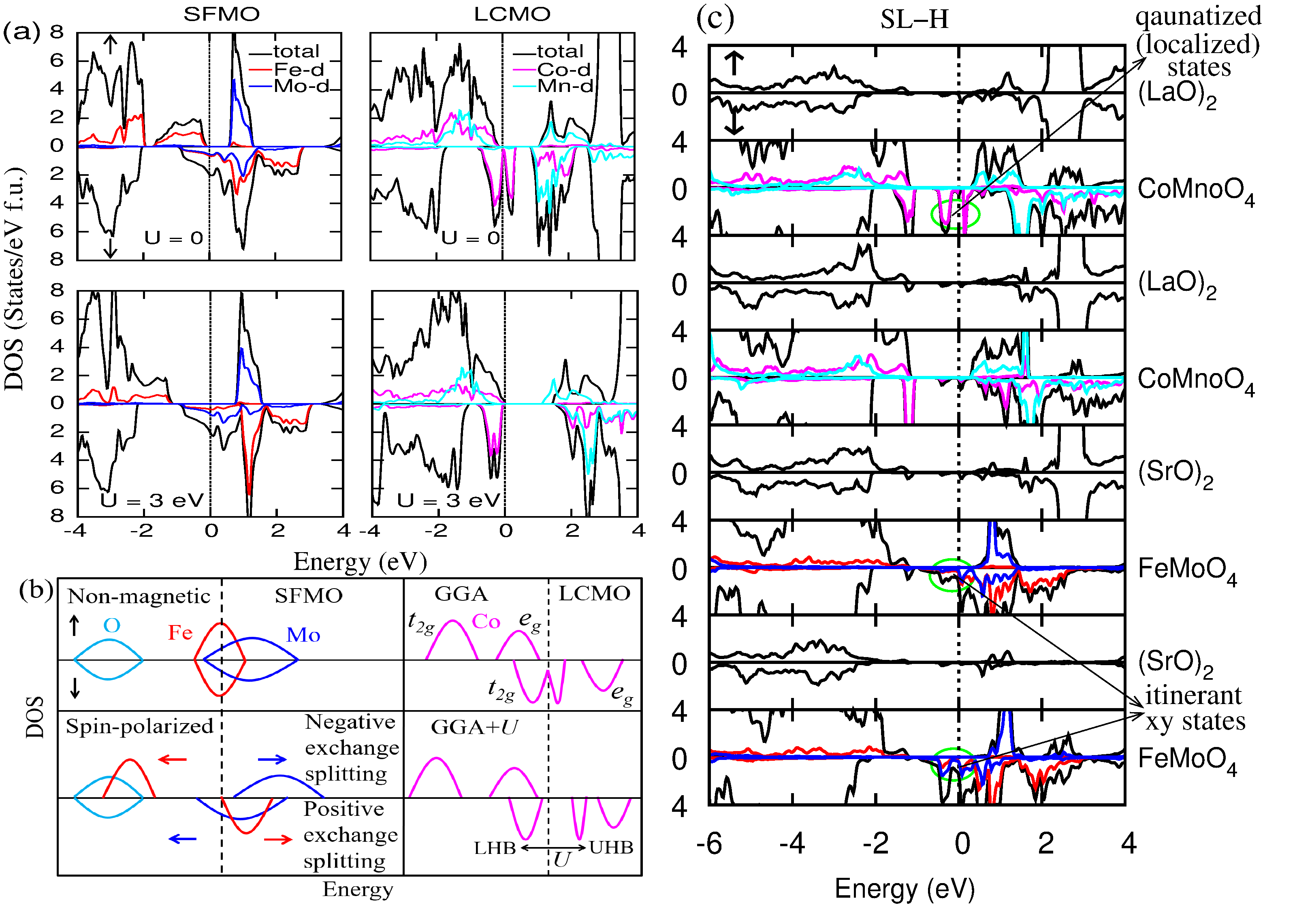}
\caption{(a) Spin resolved total and partial densities of states of bulk SFMO and LCMO. For SFMO, the gap in the the spin-up channel exists even without the strong correlation effect. The partially occupied dispersive Fe-Mo $t_{2g}$ hybridized bonding state makes the spin-down channel conducting. Within GGA, ferromagnetic LCMO is insulating in spin-up channel, and it exhibits a pseudogap at E$_F$ in spin-down channel. With inclusion of strong onsite correlation, the real gap opens up to make the system insulating. (b) The schematic illustration of the mechanism that makes SFMO half-metallic and LCMO insulating. (c) Layer resolved total and partial DOS for (SFMO)$_2$/(LCMO)$_2$ SL-H superlattice. These layer resolved DOS replicate the bottom panel of Fig. \ref{fig:Structure_SL_H_L_mechanism}(c).}
\label{fig:bulk_sfmo_lcmo_total_dos_u0}
\end{center}
\end{figure}

The spin-resolved total and partial DOS of Fig. \ref{fig:bulk_sfmo_lcmo_total_dos_u0}(a), obtained within DFT and DFT+$U$, describe the bulk electronic structure of Sr$_2$FeMoO$_6$ and La$_2$CoMnO$_6$. Fig. \ref{fig:bulk_sfmo_lcmo_total_dos_u0}(b) schematically illustrates the mechanism responsible for half-metallic behavior of Sr$_2$FeMoO$_6$ and insulating behavior of La$_2$CoMnO$_6$.

In a non-magnetic configuration, half-filled 3$d$ states of Fe$^{3+}$ and partially occupied 4$d$ states of Mo$^{5+}$ create a high DOS at E$_F$ to make the system unstable and the system becomes stable through spin-polarization. Now with large spin-exchange split, the Fe-$d$ states are occupied in the spin-up channel and are empty in the spin-down channel. However, there is a larger overlap between the Fe-$t_{2g}$ and Mo-$t_{2g}$ spin-down bands, which leads to a stronger hybridization between these two, where the bonding band is more predominantly occupied by the Mo-$t_{2g}$ states. This creates a sort of negative exchange splitting \cite{Saitoh2002,Nanda2006}.

The compound LCMO is a correlated insulator. In absence of any lattice distortion and strong onsite correlation effect, LCMO is half-metallic within GGA. Due to the distorted CoO$_6$ octahedra, there is a partial removal of the three-fold degeneracy of the $t_{2g}$ states which gives rise to a pseudogap at E$_F$ \cite{Parida2017}. However, with inclusion of onsite correlation $U$, a gap is opened up in the spin-down channel by splitting the Co-$t_{2g}$ state into lower Hubbard band (LHB) and upper Hubbard band (UHB). 

With the formation of the superlattice, following the steps proposed in Fig. \ref{fig:Structure_SL_H_L_mechanism}(c), the bulk $xz$ and $yz$ states are quantized leaving the partially occupied dispersive spin-down $xy$ states of SFMO unchanged and finally form the 2DEG. This reflected in the superlattice DOS plotted in \ref{fig:bulk_sfmo_lcmo_total_dos_u0}(c).

\vspace*{-0.4cm}

\section{Effect of strain on bulk compounds}

\begin{figure}[H]
\begin{center}
\includegraphics[angle=-0,origin=c,height=5.0cm,width=8.6cm]{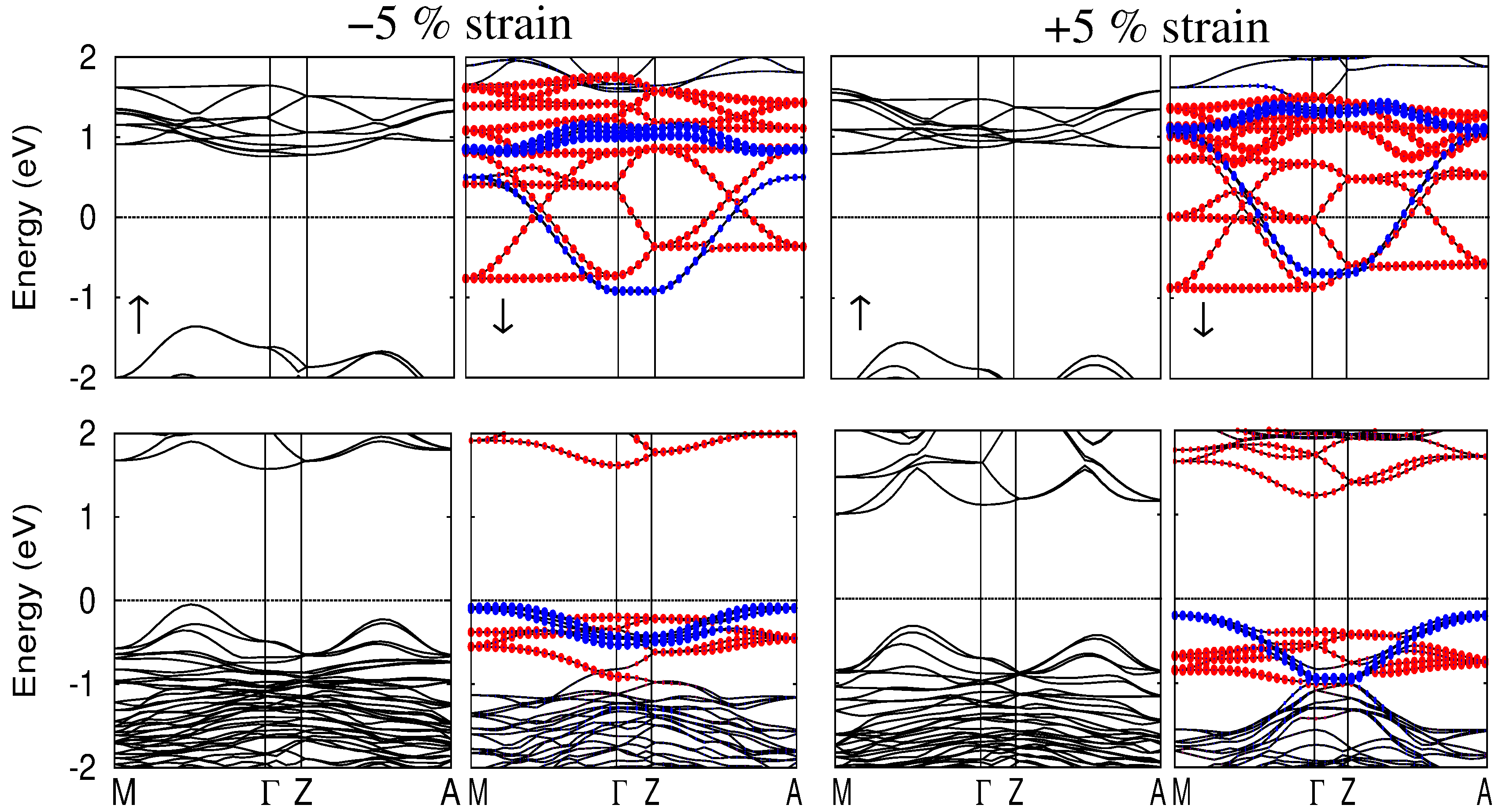}
\caption{(GGA+$U$; $U$ = 3 eV) Spin-polarized band structure of bulk SFMO (top panel) and LCMO (bottom panel) in the presence of compressive and tensile strain. The results are obtained using a four formula unit cell. The color code is same as in Fig. \ref{fig:SL-H_u0}}
\label{fig:Strain_effect}
\end{center}
\end{figure}

\vspace*{-0.4cm}

Superlattices grown on substrates experience epitaxial strain which can influence their electronic structure. In this regard examining the strain effect on the respective bulk compounds is a good starting point. In Fig. \ref{fig:Strain_effect}, we have shown the bulk band structure of SFMO and LCMO in the presence of $\pm$5 \% epitaxial strain. We find that irrespective of the nature of strain, compressive or tensile, both of the compounds are insulating in the spin up channel akin to the unstrained condition. The spin-down channel, which is responsible for the eigenstate reconstruction to create the 2DEG, also retained the metallic and insulating behavior of SFMO and LCMO respectively. In fact, there is no change in the shape of band dispersion, except for a minor variation in the band width. Hence, it can be inferred that in the absence of extreme strain condition, the 2DEG formation in the (SFMO)$_2$/(LCMO)$_2$ superlattice will remain invariant.

\section{Invariance of bulk and superlattice electronic structure with different $U$}

\begin{figure}[H]
\begin{center}
\includegraphics[angle=-0,origin=c,height=10.0cm,width=8.5cm]{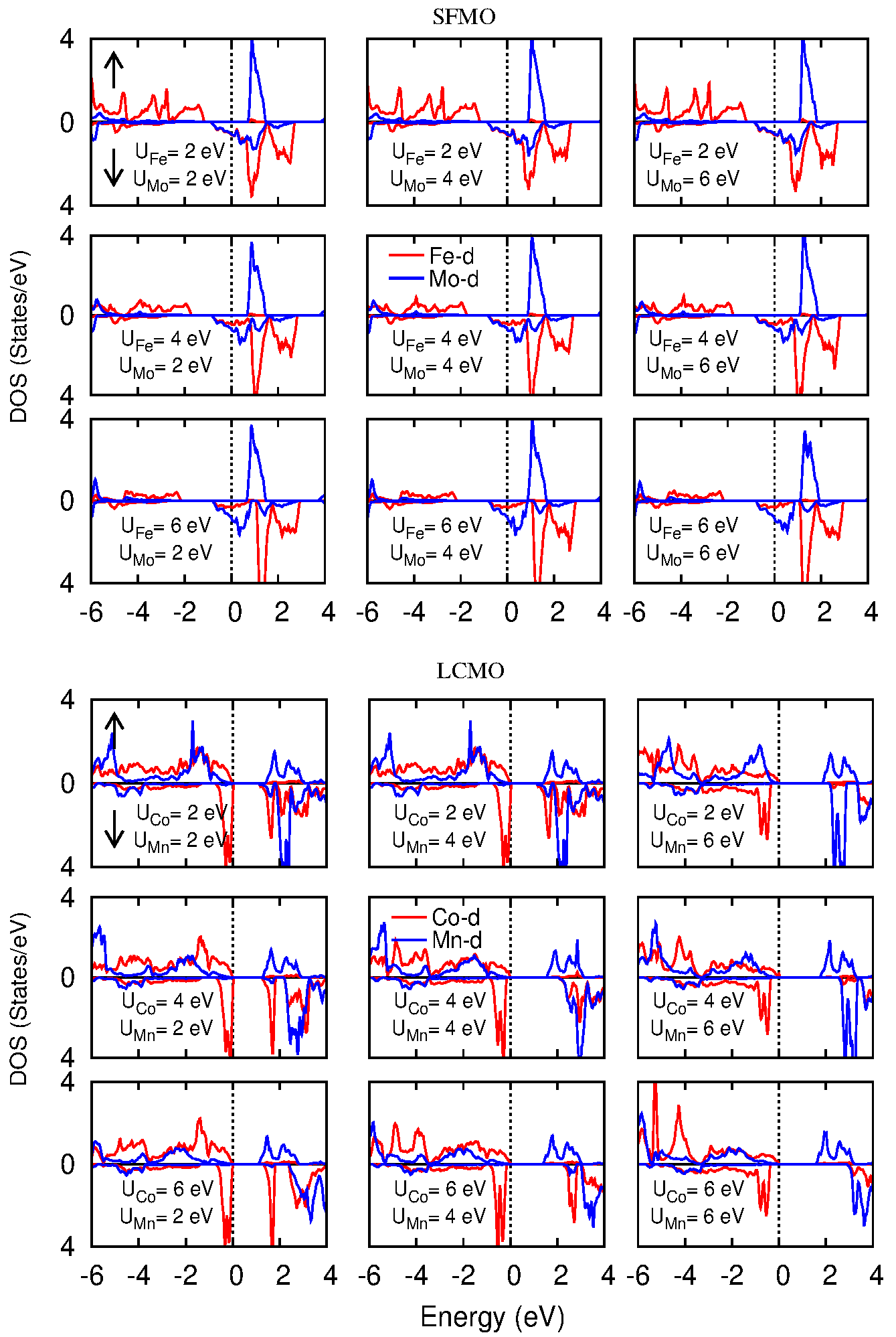}
\caption{GGA+$U$ spin-polarized partial densities of states for bulk SFMO (upper panel) and LCMO (lower panel). Different $U$ values have been considered for different transition metal elements. Irrespective of $U$ values, SFMO and LCMO retain their half-metallic and insulating behavior respectively.}
\label{fig:sfmo_lcmo_Us}
\end{center}
\end{figure}

\begin{figure}[H]
\begin{center}
\includegraphics[angle=-0,origin=c,height=4.0cm,width=8.5cm]{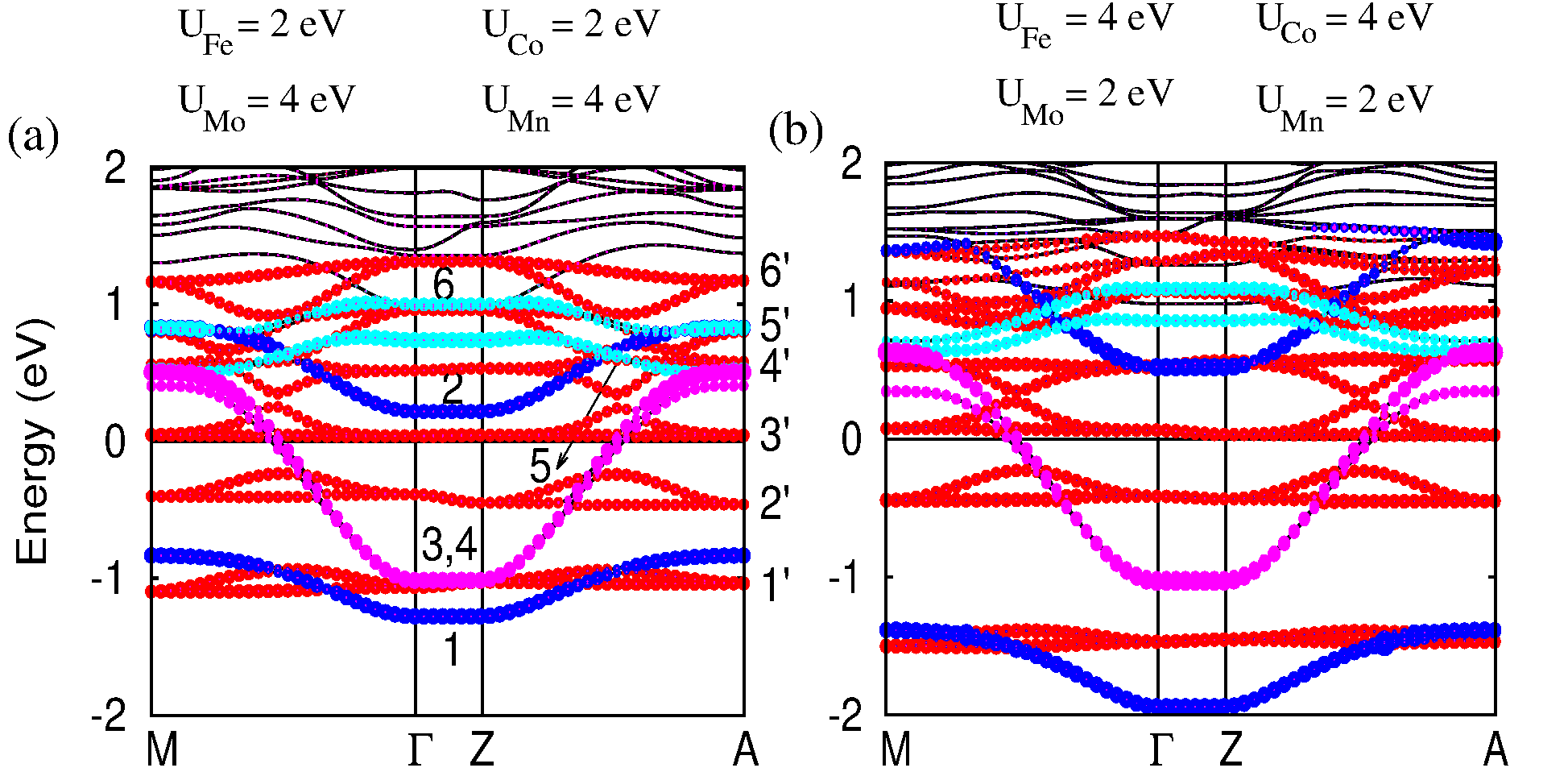}
\caption{Spin-down band structure of SL-H superlattice obtained using two different set of $U$ values as indicated in the figure. Here, we infer that the formation of 2DEG is an invariant phenomena in this superlattice. The color code is same as in Fig. \ref{fig:SL-H_u0}.}
\label{fig:SL_H_Us}
\end{center}
\end{figure}

Since both LCMO and SFMO are strongly correlated oxides, it is expected that the onsite repulsion $U$ has a major role in determining their ground state electronic structure. In the main text, we have discussed the results with same value of $U$ (= 3 and 5 eV) for Co, Mn, Fe, and Mo.  However, in general $U$ is different for different element. In this section, we report whether the salient features of the bulk and superlattice electronic structure of these double perovskites remain the same even if different $U$ values are used. In Fig. \ref{fig:sfmo_lcmo_Us}, we halve plotted the Fe-$d$ and Mo-$d$ DOS for SFMO and Mn-$d$ and Co-$d$ DOS for LCMO for different pair of $U$ values for each of them. We find that half-metallic nature of SFMO and insulating nature of LCMO are not affected. However, there is a minor redistribution of the states around the Fermi level. Such redistribution brings minor change in the occupancy of the $d$-states for the half-metallic system. The spin-down band structures plotted in Fig. \ref{fig:SL_H_Us} suggest that the formation of 2DEG in the (SFMO)$_2$/(LCMO)$_2$ superlattice as well as quantization of the states, as discussed in the main text, are universal and are not affected by the change in $U$ values. However, as expected the band centers of the lower and upper Hubbard bands vary with $U$.

\vspace*{-1.2cm}

\section{Effect of superlattice period}

\vspace*{-0.4cm}

\begin{figure}[H]
\begin{center}
\includegraphics[angle=-0,origin=c,height=4.0cm,width=8.5cm]{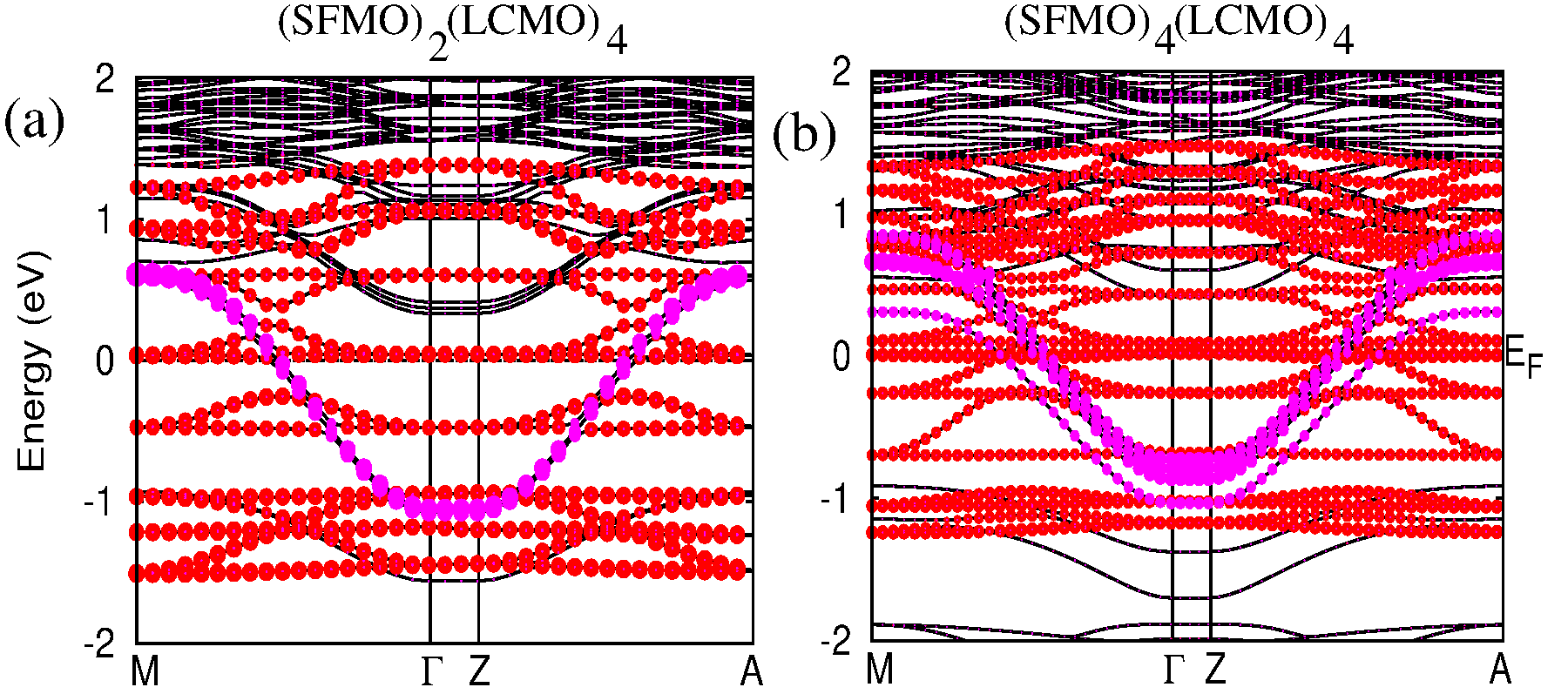}
\caption{The spin-down band structure of (2,4) and (4,4) SFMO/LCMO superlattices. The calculations are carried out with $U$ = 3 eV on a relaxed structure. The color code is same as in Fig. \ref{fig:SL-H_u0}.}
\label{fig:SL_period}
\end{center}
\end{figure}

\vspace*{-0.4cm}

As the eigenstate reconstruction of this superlattice depends on the potential profile of the system (see Fig. \ref{fig:avg_potential}), it is expected that the period of the superlattice will have influence on its electronic structure. With this as objective, in Fig. \ref{fig:SL_period}, we have plotted the band structure of (SFMO)$_2$/(LCMO)$_4$ and (SFMO)$_4$/(LCMO)$_4$ superlattices. We find that the itinerant behavior of the $xy$ dominated bands is not affected with the superlattice period. However, though $xz$ and $yz$ dominated bands are discretized, the extent of localization of these states is affected with the thickness of SFMO. With increasing thickness, the electron localization along the superlattice growth direction reduces which weakens the formation of spin-polarized 2DEG. However, this is a preliminary study and a detailed in-depth study is required to have better understanding on the effect of superlattice period on the electronic structure.

\bibliography{Paper}

\end{document}